# Magnetic nanoparticle detection based on nonlinear Faraday rotation

Xinchao Cui[1], Feidi Xiang[1], Chong Lu[2], Chunping Liu[2], Wenzhong Liu[3, 4, a]

**AFFILIATIONS**

1 School of Artificial Intelligence and Automation, Huazhong University of Science and Technology, Wuhan 430074, China

2 Department of Thyroid and Breast Surgery, Union Hospital, Tongji Medical College, Huazhong University of Science and Technology, Wuhan 430022, China

3 China-Belt and Road Joint Laboratory on Measurement and Control Technology, Huazhong University of Science and Technology, Wuhan 430074, China

4 Shenzhen Huazhong University of Science and Technology Research Institute, Shenzhen 518063, China

a) E-mail: lwz7410@hust.edu.cn

**ABSTRACT**

Magnetic nanoparticle (MNP) has attracted interest in various research fields due to its special superparamagnetic and strong magneto-optical effects, especially as contrast agents to enhance the contrast of medical imaging. By introducing the interaction coefficient, we propose a model of the nonlinear Faraday rotation of MNP under the excitation of an external alternating magnetic field. In our homemade device (which can detect the rotation angle as low as about 2e-7 rad), it has been verified that the higher harmonics of the Faraday rotation can avoid the interference of paramagnetic and diamagnetic background at lower concentrations. What's more, the higher harmonics of the Faraday rotation of MNP can be detected in real time and they have a linear relationship with concentration. In the future, it is expected to use MNP as a magneto-optical contrast agent to achieve high-resolution imaging in vivo.

Magnetic nanoparticle (MNP) has been widely studied for its excellent and diverse physical properties. Especially in biomedical applications, MNP is often used as contrast agents or tracers after surface modification to complete biological tissues imaging in vivo[1-4]. Among them, magnetic resonance imaging (MRI) and magnetic particle imaging (MPI) techniques are representative[5-11]. As the contrast agent, MNP affects the relaxation time of the echo signal and effectively improves the contrast of images in MRI[5,6]. In addition, MPI is considered one of the most promising imaging techniques for metabolic imaging, as the spatial concentration distribution of MNP is directly inverted by measuring the magnetization response of MNP[7-9]. However, the full width at half maxima (FWHM) of the point spread function (PSF) in MRI and MPI is limited by the spatially encoded magnetic field. Therefore, it is difficult to break through the theoretical limit of 1 mm in the resolution of these imaging methods[10,11].

Optical detection of MNP is considered to be one of the most promising technologies to break through the bottleneck of high-resolution imaging in vivo. For example, it is possible to image microstructures such as brain vasculature and nerves in near-infrared biological window (700–900 nm) because the resolution is only a few μm[12-15]. MNP suspension has been verified to have strong magneto-optical effects[16-18], and its magnetic-optical film can not only change the transmittance of the incident light[19,20], but also produce Faraday magneto-optical effect[21,22] and Cotton-Mouton effect[23,24] under the excitation of an external magnetic field. Therefore, MNP can be used as a magneto-optical contrast agent for high-resolution optical imaging of tissues in vivo.

The most important feature of MNP as a magneto-optical contrast agent is that the polarization state of light is independent of the light intensity during propagation, i.e., the attenuation of the light intensity does not affect the angle of the polarization plane. Although light is severely absorbed and scattered as it propagates through human tissues such as skin and fat[25], the interference of the attenuation of light intensity can be avoided by extracting the polarization angle. The concentration detection or imaging of MNP can be achieved by using weak light detection techniques such as single photon counters in the future. Most previous studies on the magneto-optical effects of MNP have used shorter light paths and higher concentrations, with the MNP suspension film equated to an effective medium[26-28]. The concentration of MNP as a contrast agent in vivo imaging is limited by clinical standards[8]. At this point, the Faraday rotation produced by antimagnetic or paramagnetic backgrounds such as water and other media would not be negligible and the effective medium theory would no longer be appropriate.

In this letter, we propose an improved model based on the Langevin equation that successfully explains the nonlinear Faraday rotation produced by the special superparamagnetic magnetization response of the MNP under an external magnetic field. The Faraday rotation produced by the MNP under a sinusoidal alternating magnetic field of frequency $f$ is decomposed into odd harmonics such as $f, 3f, 5f$ by Fourier series expansion. In contrast to the fundamental harmonic $f$, the detection of higher harmonics is not affected by the antimagnetic or paramagnetic background. The validity of the nonlinear magneto-optical effect model of MNP was demonstrated experimentally using a homemade magneto-optical detection device. The experimental results show that the real-time quantitative detection of the higher harmonics of Faraday rotation can be realized by eliminating antimagnetic and paramagnetic background interference in principle. This provides a strong support for the use of MNP as a magneto-optical contrast agent for high-resolution imaging in vivo.

The Faraday magneto-optical effect can be explained as the difference in the refractive index of the medium for the left- and right-handed circularly polarized light under an external magnetic field. Therefore, after the linearly polarized light propagates a certain distance inside the medium, the phase difference of the left and right handed circularly polarized light changes. This causes the polarization plane of the transmitted light to rotate by an angle relative to the incident light. The angle changed in the polarization plane is generally defined as the Faraday rotation angle[29], which is written as

$$\theta = BVL \quad (1)$$

where $B$ is the magnetic flux density of external magnetic in the direction of light propagation within the medium. Under the excitation of an external magnetic field, the magnetic permeability of paramagnetic and antimagnetic media is almost close to 1, so the magnitude of their magnetic flux density can be approximated as the external magnetic field. $V$ is the Verdet constant of the medium, which is related to the wavelength of the incident light $\lambda$ and the refractive index of the medium for left and right handed circularly polarized light $n_+, n_-$, and is consistent with $V = \pi (n_+ - n_-)/\lambda$. $L$ is the path of light passing through the medium along the direction of the external magnetic field. MNP is superparamagnetic and enhances the magnetization response of the medium under the external field, so the MNP suspension is often used to make thin film to enhance the Faraday rotation. Therefore, the Faraday rotation angle of the MNP suspension defines the sum of the magnetization response of the medium itself and that the MNP, which is approximated as

$$\theta = (H + vM)VL \quad (2)$$

where $H$ is the external magnetic field, $v$ is the interaction coefficient, and $M$ is the magnetization response of MNP, which conforms to the Langevin equation $L(x) = \coth(x) - 1/x$. The magnetization response of the MNP suspension can be inverted by detecting the Faraday rotation angle. During light propagation, the polarization state is determined by the phase and is not affected by the light intensity. Therefore, using coherent detection can enable real-time, remote, non-contact detection of MNP.

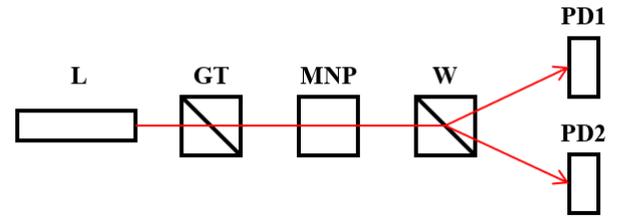

**Fig. 1 The schematic of Faraday rotation detection device. L is a helium-neon laser; GT is a Glenn Taylor prism; MNP is the sample; W is a Wollaston prism; PD1, PD2 are the differential inputs of the balance detector. The angle between the axis of the Wollaston prism and the Taylor prism is π/4.**

The experiments were carried out using a homemade Faraday rotation detection device, and the schematic is shown in Fig. 1. A helium-neon laser with a wavelength of 632.8 nm (HRS015, Thorlabs) was used as the light source to generate linearly polarized light. The light was first passed through a Glenn Taylor prism to produce an extinction ratio of better than $10^5$:1, and then passed through the MNP suspension. A linear power amplifier (7224, AE Tech) is driven by the data acquisition card (PXIe 6396, NI) provided energy to solenoid to generate a sinusoidal alternating magnetic field excitation at frequency of $f = 110$ Hz. Then the transmitted light is divided into two beams orthogonal in the polarization plane through the Wollaston prism, and finally measured using a balance detector (PDB210A, Thorlabs). The sensitivity of detection is maximized when the angle between the axis of the polarizer and the analyzer is set to π/4[30]. According to Marius's theorem, the light intensity information received by the two differential inputs of the balance detector can be expressed as

$$I_1 = I_0 \sin^2(\theta + \pi/4)$$
$$I_2 = I_0 \cos^2(\theta + \pi/4) \quad (3)$$

where $I_0$ is the intensity of light transmitted through the MNP suspension. Simplified by trigonometric formulas, it can be calculated that the differential signal of the balance detector is $I_1 - I_2 = I_0 \sin(2\theta)$, and the common signal is $I_1 + I_2 = I_0$. When the Faraday rotation angle of the MNP suspension is smaller, there is

$$\theta \approx \frac{\sin(2\theta)}{2} = \frac{I_1 - I_2}{2(I_1 + I_2)} \quad (4)$$

The interference of the light intensity $I_0$ can be eliminated by Eq. (4), thus realizing the detection of Faraday rotation angle in real time. Therefore, the dynamic response to MNP can be achieved under the excitation of the alternating magnetic field $H(t) = H_0 \sin(2\pi f t)$. The magnetization response of the MNP is nonlinear, and it is usually expressed as a sum of odd harmonics by Fourier expansion, denoted as

$$M(t) = \sum_{\substack{n=1,\\k=2n-1}}^{\infty} \alpha_k M_k \sin(k 2\pi f t + \phi_k) \quad (5)$$

where $M_k$ is the amplitude of the $k$ th harmonic of the magnetization response, $\alpha_k = 1/\sqrt{1 + (k 2\pi f \tau)^2}$ and $\varphi_k = \text{atan}(k 2\pi f \tau)$ are the amplitude attenuation and phase delay of each harmonic caused by the dynamic relaxation. Where $\tau$ is the relaxation time of MNP, which is less than 1e-5 s in our experiments. So $\alpha_k$ and $\phi_k$ can be ignored at lower excitation frequency, i.e. the magnetization response of MNP is superparamagnetic. Bringing $H(t)$ and $M(t)$ into Eq. (2) shows that the relationship between Faraday rotational angle and the MNP magnetization response under lower frequency alternating magnetic field excitation can be obtained by

$$\theta(t) = H_0 V L \sin(2\pi f t) + \sum_{\substack{n=1,\\k=2n-1}}^{\infty} v M_k V L \sin(k 2\pi f t) \quad (6)$$

Therefore, the magnetization response of MNP under the alternating magnetic field excitation can be detected by the Faraday rotational angle. Among them, the fundamental harmonic of $\theta(t)$ contains not only the response of MNP but also the paramagnetic and diamagnetic background media. But the higher harmonics are only correlated and proportional to the magnetization response of MNP.

The particles used for the experiments were commercially purchased (SHP-30, Ocean NanoTech) with an iron concentration of 5 mg/mL. They consist of $Fe_3O_4$ core and carboxylic acid surface modification layer with a core size of 30 nm and a hydrodynamic diameter of 32.67 nm[31]. In our experiments, MNP was diluted with ultra-pure water to concentrations of 0, 50, 100, 250, and 500 μg/mL. Then the suspension was held in a 10*10*25 mm³ quartz cuvette to ensure good transmittance to the incident light. As is shown in Fig. 2, It can be seen that the color of MNP gradually becomes darker and the intensity of transmitted light gradually decreases with the increase of concentration. The intensity of transmitted light decays with increasing concentration almost in accordance with the function $a \exp(bx)$. MNP concentrations above 500 μg/mL in the 10 mm optical path severely attenuate the intensity of transmitted light, and a photodetector with higher conversion efficiency is needed for the measurement.

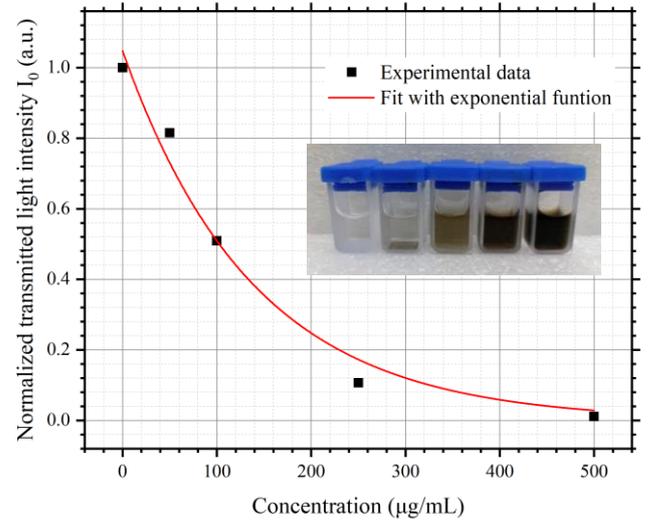

**Fig. 2 Experimental results of the transmitted light intensity vs. MNP concentration.**

Paramagnetic and antimagnetic media (including water and quartz cuvettes) produce an additional smaller Faraday rotation under the external magnetic field[22,32], which can be neglected at higher concentrations of MNP. However, as the MNP concentration decreases, the volume fraction decreases and this angle is sufficient to affect the detection of the MNP magneto-optical effect. The fundamental harmonic of Faraday rotation for the experimental results of MNP at different concentrations are shown in Fig. 3. The magnetic field excitation with frequency $f$ = 110 Hz and amplitude from 0 to 436 Oe. It can be seen that the Faraday rotation of the suspension with MNP concentration of 50 μg/mL (100 times dilution) is too small and close to the result for water. The phase of the fundamental harmonic at different concentrations in the Fig. 3 is almost the same as that of water, i.e., the direction of the Faraday rotation produced by MNP is the same as that of water. Although the magnetization response of MNP is much higher than that of water, the Faraday rotation is not enhanced by orders of magnitude. Therefore, we consider that MNP only enhances the local magnetic field around the water by interaction, and the effect can be expressed by the interaction coefficient $v$.

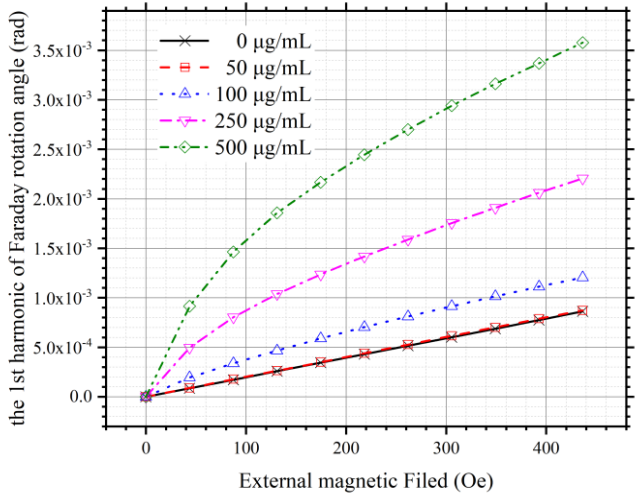

**Fig. 3 Experimental results of the fundamental harmonic of Faraday rotation angle vs. external magnetic field with different MNP concentrations. The concentration 0 μg/mL is water.**

According to the proposed theoretical Eq. (6), it is known that the magnetization response of background produces the Faraday rotation only with the fundamental harmonic and without the higher harmonics under an external alternating magnetic field. In contrast, the magnetization response of MNP produces and is proportional to the higher harmonics of Faraday rotation. Therefore, the MNP can be specifically detected by the high harmonics of the Faraday rotation. As shown in Fig. 4, the nonlinear magneto-optical effect of MNP produces 3rd-9th harmonics that can be accurately detected for different concentrations, while water has no signal. The Faraday rotation produced by MNP increases gradually with the increases of external magnetic field, but the slope decreases. And the amplitude of each higher harmonic decreases with increasing order, which is consistent with the results of the magnetization response of MNP[33]. In particular, as shown in Fig. 4(d), the nonlinear Faraday rotation of MNP at a concentration of 50 μg/mL is very small. This is because the volume fraction of MNP in the suspension is much smaller than that of the water and the interaction is very weak. Our device can achieve minimum rotation angle detection of about 2e-7 rad subject to interference form $1/f$ noise and shot noise. Therefore, it is currently not possible to accurately detect the Faraday rotation angle at higher concentrations.

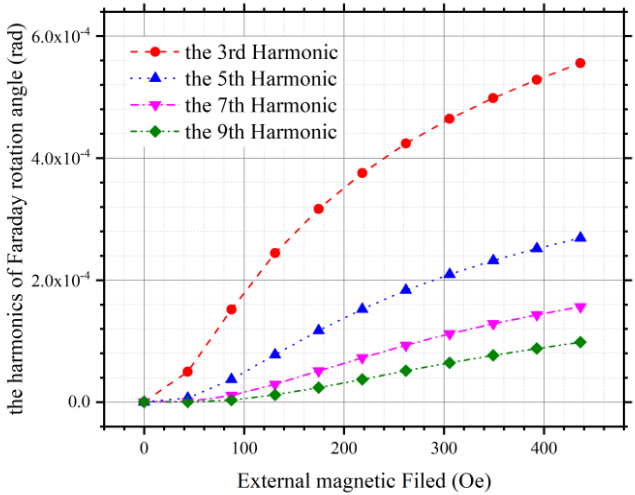

(a)

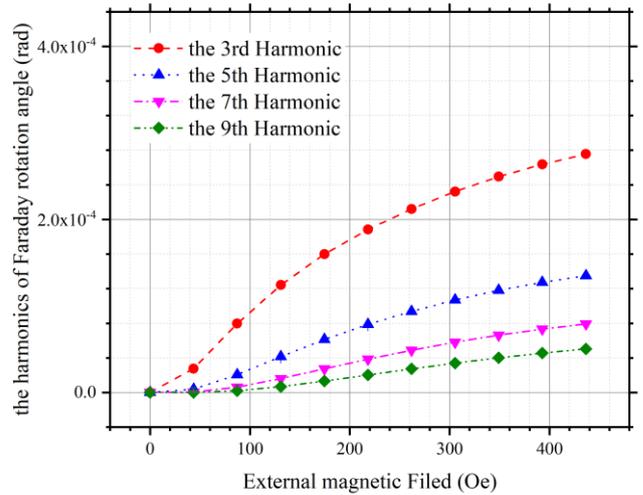

(b)

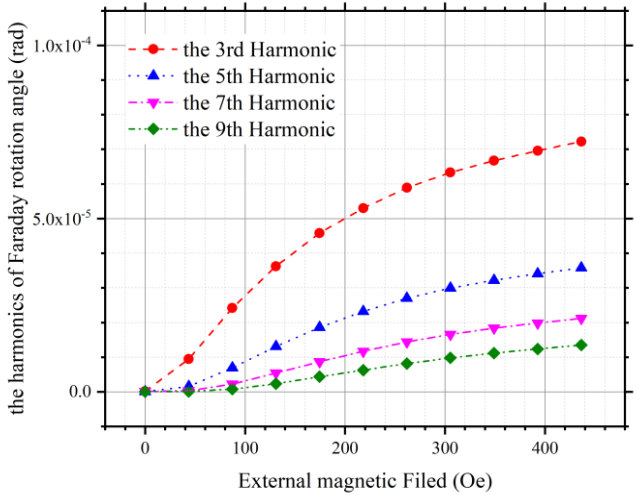
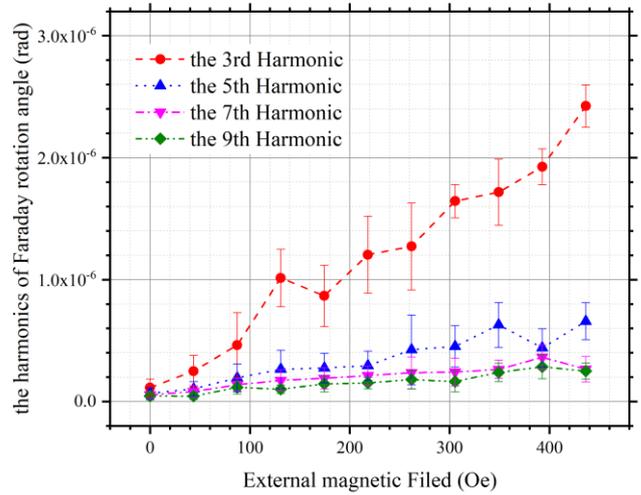

(c)  (d)

Fig. 4 Experimental results of the 3rd-9th harmonics of Faraday rotation angle vs. external magnetic field with different MNP concentrations. The concentrations are (a) 500 μg/mL, (b) 250 μg/mL, (c) 100 μg/mL and (d) 50 μg/mL.

In order to visually compare the higher harmonics of Faraday rotation with different concentrations of MNP under external magnetic field, the different harmonics for 436 Oe magnetic field are plotted in Fig. 5. It can be seen that the harmonics almost linearly increase with the concentrations of MNP. The MNP is uniformly dispersed with the water in suspension and the interaction coefficient $v$ can be considered as concentration independent. Thus, quantitative detection of MNP can be achieved by the higher harmonics of the Faraday rotation, which offers the possibility of using it as contrast agents for magneto-optical imaging.

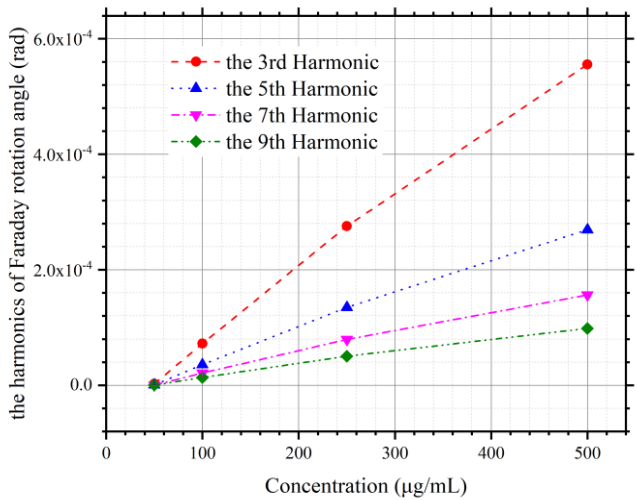

Fig. 5 Experimental results of the harmonics of Faraday rotation angle vs. MNP concentration under 436 Oe external magnetic field.

Our proposed model can reasonably explain the nonlinear Faraday rotation of MNP suspensions under low-frequency alternating magnetic field excitation using interaction coefficients, and has been verified by a homemade device. The angle of polarization is independent of attenuation of light intensity at higher concentrations, but the signal-to-noise ratio can be improved by increasing the light intensity appropriately. What's more, the penetration of light and the resolution limit of imaging depend on the wavelength of light. Since the interference of the background on the Faraday rotation at lower concentrations can be directly avoided by using higher harmonics, MNP is very promising as a magneto-optical contrast agent for high-resolution imaging in vivo.

This work was supported by the National Key R&D Program of China (Grant Nos. 2022YFE0107500, 2022YFE0204700), the National Natural Science Foundation of China (Grant No. 61973132), and Shenzhen Municipal Scientific Program (Grant Nos. JCYJ20200109110612375, JCYJ20210324142004012).

**REFERENCE**
[1] Maria Hepel, Magnetochemistry **6** (1) (2020).
[2] K. Wu, D. Q. Su, R. Saha, J. M. Liu, V. K. Chugh, and J. P. Wang, Acs Appl Nano Mater **3** (6), 4972 (2020).
[3] L. C. Wu, Y. Zhang, G. Steinberg, H. Qu, S. Huang, M. Cheng, T. Bliss, F. Du, J. Rao, G. Song, L. Pisani, T. Doyle, S. Conolly, K. Krishnan, G. Grant, and M. Wintermark, AJNR Am J Neuroradiol **40** (2), 206 (2019).
[4] Nahid Rezvani Jalal, Parvaneh Mehrbod, Shahla Shojaei, Hagar Ibrahim Labouta, Pooneh Mokarram, Abbas Afkhami, Tayyebeh Madrakian, Marek J. Los, Dedmer Schaafsma, Michael Giersig, Mazaher Ahmadi, and Saeid Ghavami, Acs Appl Nano Mater **4** (5), 4307 (2021).
[5] M. J. Chen, S. H. Liao, H. C. Yang, H. Y. Lee, Y. J. Liu, H. H.

Chen, H. E. Horng, and S. Y. Yang, Journal of Applied Physics **110** (12) (2011).

[6]J. H. Hankiewicz, Z. Celinski, K. F. Stupic, N. R. Anderson, and R. E. Camley, Nat Commun **7**, 12415 (2016).

[7]J. Rahmer, J. Weizenecker, B. Gleich, and J. Borgert, BMC Med Imaging **9**, 4 (2009).

[8]J. Weizenecker, B. Gleich, J. Rahmer, H. Dahnke, and J. Borgert, Phys Med Biol **54** (5), L1 (2009).

[9]B. Gleich and J. Weizenecker, Nature **435** (7046), 1214 (2005).

[10]C. Billings, M. Langley, G. Warrington, F. Mashali, and J. A. Johnson, Int J Mol Sci **22** (14) (2021).

[11]P. W. Goodwill and S. M. Conolly, IEEE Trans Med Imaging **29** (11), 1851 (2010).

[12]Z. Feng, X. Yu, M. Jiang, L. Zhu, Y. Zhang, W. Yang, W. Xi, G. Li, and **J. Qian**, Theranostics **9** (19), 5706 (2019).

[13]Isabella M. Kopton and Peter Kenning, Frontiers in Human Neuroscience **8** (2014).

[14]J. Qi, C. Sun, A. Zebibula, H. Zhang, R. T. K. Kwok, X. Zhao, W. Xi, J. W. Y. Lam, J. Qian, and B. Z. Tang, Adv Mater **30** (12), e1706856 (2018).

[15]Felix Scholkmann, Stefan Kleiser, Andreas Jaakko Metz, Raphael Zimmermann, Juan Mata Pavia, Ursula Wolf, and Martin Wolf, NeuroImage **85**, 6 (2014).

[16]John Philip and Junaid M. Laskar, Journal of Nanofluids **1** (1), 3 (2012).

[17]H. W. Davies and J. P. Llewellyn, Journal of Physics D: Applied Physics **13** (12), 2327 (1980).

[18]Susamu Taketomi, Masakazu Ukita, Masaki Mizukami, Hideki Miyajima, and Soshin Chikazumi, Journal of the Physical Society of Japan **56** (9), 3362 (1987).

[19]J. M. Laskar, J. Philip, and B. Raj, Phys Rev E Stat Nonlin Soft Matter Phys **78** (3 Pt 1), 031404 (2008).

[20]Dillip Kumar Mohapatra, Junaid Masud Laskar, and John Philip, Journal of Molecular Liquids **304** (2020).

[21]R. J. Murdock, S. A. Putnam, S. Das, A. Gupta, E. D. Chase, and S. Seal, Small **13** (12) (2017).

[22]R. Soucaille, M. E. Sharifabad, N. D. Telling, and R. J. Hicken, Appl Phys Lett **116** (6) (2020).

[23]K. L. Chen, Z. Y. Yang, and C. W. Lin, J Nanobiotechnology **19** (1), 301 (2021).

[24]B. Y. Ku, M. L. Chan, Z. Ma, and D. A. Horsley, J Magn Magn Mater **320** (18), 2279 (2008).

[25]Frans F. Jöbsis, Science **198** (4323), 1264 (1977).

[26]Ondřej Vlašín, Oana Pascu, Anna Roig, and Gervasi Herranz, Physical Review Applied **2** (5) (2014).

[27]Yu A. Barnakov, B. L. Scott, V. Golub, L. Kelly, V. Reddy, and K. L. Stokes, Journal of Physics and Chemistry of Solids **65** (5), 1005 (2004).

[28]P. M. Hui and D. Stroud, Appl Phys Lett **50** (15), 950 (1987).

[29]P. S. Pershan, Journal of Applied Physics **38** (3), 1482 (1967).

[30]Nihad A. Yusuf, Akram A. Rousan, and Hassan M. El-Ghanem, Journal of Magnetism and Magnetic Materials **65** (2-3), 282 (1987).

[31]Xinchao Cui, Chong Lu, Chunping Liu, and Wenzhong Liu, Sensors and Actuators A: Physical **357** (2023).

[32]D. L. Carr, N. L. R. Spong, I. G. Hughes, and C. S. Adams, Eur J Phys **41** (2) (2020).

[33]Xinchao Cui, Lan Li, and Wenzhong Liu, IEEE Transactions on Instrumentation and Measurement **71**, 1 (2022).